\documentclass[aps,prb,twocolumn,showpacs]{revtex4}
\usepackage{amssymb}
\usepackage{graphicx}

\begin{document}

\title{Impurity effect and suppression to superconductivity in Na(Fe$_{0.97-x}$Co$_{0.03}$T$_x$)As (T=Cu, Mn)}

\author{Qiang Deng, Xiaxin Ding, Sheng Li, Jian Tao, Huan Yang, Hai-Hu Wen$^{\star}$}

\affiliation{Center for Superconducting Physics and Materials,
National Laboratory of Solid State Microstructures and Department
of Physics, Nanjing University, Nanjing 210093, China}

\begin{abstract}

We report the successful growth and the impurity scattering effect of single
crystals of Na(Fe$_{0.97-x}$Co$_{0.03}$T$_x$)As (T=Cu, Mn). The temperature dependence of DC magnetization
at high magnetic fields is measured for different concentrations of Cu and Mn. Detailed analysis based on the Curie-Weiss law indicates that the Cu doping weakens the average magnetic moments, while doping Mn enhances the local magnetic moments greatly, suggesting that the former may be non- or very weak magnetic impurities, and the latter give rise to magnetic impurities. However, it is found that
both doping Cu and Mn will enhance the residual resistivity
and suppress the superconductivity at the same rate in the
low doping region, being consistent with the prediction of
the S$^{\pm}$ model. For the Cu-doped system, the superconductivity
is suppressed completely at a residual resistivity $\rho_0$ = 0.87 m$\Omega$ cm at which a strong localization effect is observed. However,
in the case of Mn doping, the behavior of suppression to \emph{T}$_{c}$
changes from a fast speed to a slow one and keeps superconductive even up to a residual resistivity of 2.86 m$\Omega$ cm.
Clearly the magnetic Mn impurities are even not as detrimental as the
non- or very weak magnetic Cu impurities to superconductivity in the high doping regime.
\end{abstract}

\pacs{74.20.Rp, 74.70.Dd, 74.62.Dh, 74.70.Xa} \maketitle

\section{Introduction}
The discovery of iron-based superconductors in 2008 has brought
about new vigor and vitality into the study of unconventional
superconductivity.\cite{Hosono} The pairing symmetry is very
crucial for understanding the superconducting mechanism. In the
cuprates, the pairing symmetry was proved to be
d-wave,\cite{Tsuei} while in the iron-based superconductors it is
still under debate. Theoretically, a pairing model with the gap
structure of S$^{\pm}$ was proposed. This model is based on the
assumption that the pairing is established by exchanging the
paramagnons given by the antiferromagnetic spin fluctuations
between two electrons with opposite momentum and spins. This kind
of pairing will naturally lead to a sign reversal of the order
parameter between the electron and the hole Fermi
pockets.\cite{Mazin,Kuroki,Chubukov,FWang,Hirschfeld} This model has gained
supports from many experiments, including the scanning tunneling
spectroscopy (STS) measurements\cite{Hanaguri,HYang} and inelastic
neutron scattering experiments.\cite{Christianson} On the other
hand, a pairing state without the sign reversal of the gaps,
namely S$^{++}$, was proposed if the pairing is mediated by
orbital fluctuations. \cite{Onari,Kontani,Saito} In some special
cases, even the d-wave gap may be expected in iron-based
superconductors.\cite{Kuroki,Maiti,AFWang}

In the superconducting state, the impurity scattering effect is
closely related to the gap structure, the characteristics of the
impurities and the underlying electronic structure. The impurity
scattering effect may give some important clues to unravel the
pairing gap structure as well as the pairing mechanism. According
to the Anderson's theorem,\cite{Anderson} in a conventional
s-wave superconductor, the magnetic impurity can break the Cooper
pairs easily, while superconductivity remains robustly with the
presence of nonmagnetic impurities in the unitary limit. In sharp
contrast, Anderson's theorem is seriously violated for the
unconventional superconductor, where both magnetic and nonmagnetic
impurities are detrimental to superconductivity. In cuprate
superconductors with a d-wave gap structure, significant
\emph{T}$_{c}$ suppression was observed with doping
Zn.\cite{Alloul,Julien,GXiao} In the case of S$^{\pm}$ pairing
state, it was  pointed out that both magnetic and nonmagnetic
impurities can suppress \emph{T}$_{c}$
rapidly.\cite{Onari,YYZhang,Tarantini} In order to unveil the
mystery of superconducting mechanism, plenty of experiments have
been carried out on the impurity effects in iron-based
superconductors.\cite{PCheng,PCheng2,JLi,JLi2,JLi3,Gunther,
Inabe,Hirschfeld,YLi,DTan,YFGuo, ZTZhang,Frankovsky,Kirshenbaum}
Unfortunately, the conclusions remain highly controversial.
Previous studies on Mn impurities showed that \emph{T}$_{c}$ was
strongly suppressed in
Ba$_{0.5}$K$_{0.5}$Fe$_{2}$As$_{2}$\cite{PCheng,JLi} and
Ba(Fe$_{1-y}$Co$_{y}$)$_{2}$As$_{2}$ system,\cite{JLi2} while
\emph{T}$_{c}$ is not suppressed or even enhanced in Mn-doped
FeSe$_{0.5}$Te$_{0.5}$ superconductor.\cite{Gunther} Zn impurity
suppress T$_c$ rapidly in
BaFe$_{1.89-2x}$Zn$_{2x}$Co$_{0.11}$As$_{2},$\cite{JLi3} whereas
superconducting state remains robustly in
Fe$_{1-y}$Zn$_{y}$Se$_{0.3}$Te$_{0.7}.$\cite{Inabe} Theoretically
it is proposed that the pairing symmetry in iron-based
superconductor can be different from material to
material.\cite{Hirschfeld} A study in
LaFe$_{1-y}$Zn$_{y}$AsO$_{1-x}$F$_{x}$ system showed that
superconducting transition temperature increases in the underdoped
regime with doping Zn impurities, remains unchanged in the
optimally doped regime and severely suppressed in the overdoped
regime, suggesting that the pairing symmetry could change from
S-wave to S$^{\pm}$ or even d-wave states with increasing doping
concentration.\cite{YLi} In order to have a better understanding
of the pairing symmetry and superconducting mechanism in Fe-based
superconductors, further experimental studies, especially with the
known properties of the impurities, are highly desired.

In this study, we investigate the impurity effect of doping Cu and
Mn into the 111-type iron-based superconductors
Na(Fe$_{0.97}$Co$_{0.03}$)As. Our study reveal that the Cu doping
weakens the average magnetic moments (below 0.4 $\mu_B$/Fe site),
while doping Mn enhances the local magnetic moments greatly,
suggesting that Cu dopants behave as non- or very weak magnetic
impurities, and Mn dopants are magnetic ones. It is found that
both doping Cu and Mn can enhance the residual resistivity and
suppress the superconductivity rapidly, which is consistent with
the prediction of the S$^{\pm}$ pairing model in the low doping
region.

\section{Experimental methods}
The single crystals of Na(Fe$_{0.97}$Co$_{0.03}$)As (named as pristine
sample) and  Na(Fe$_{0.97-x}$Co$_{0.03}$T$_{x}$)As (T = Cu and Mn)
 were synthesized by the self-flux method using NaAs as the flux. Firstly,
NaAs was prepared as the precursor. The Na (purity 99\%, Alfa
Aesar) was cut into pieces and mixed with As powders (purity
99.99\%, Alfa Aesar), the mixture was put into an alumina crucible
and sealed in a quartz tube in vacuum. Then it was slowly heated
up to 200$\,^{\circ}\mathrm{C}$ and held for 10 hours, followed by
cooling down to room temperature. Then the resultant NaAs, and Fe
(purity 99.9\%, Alfa Aesar), Co (purity 99.9\%, Alfa Aesar), Cu
or Mn (purity 99.9\%, Alfa Aesar) powders were weighed with an
atomic ratio of NaAs:Fe:Co:T = 4:(0.97-x):0.03:x and ground
thoroughly. The mixture was loaded into an alumina crucible, then
sealed in an iron tube under Ar atmosphere. The iron tube was then
sealed in an evacuated quartz tube to prevent the oxidization of
the iron tube. Then it was placed into the furnace and heated up
to 950$\,^{\circ}\mathrm{C}$ and held for 10 hours, followed by
cooling down to 600$\,^{\circ}\mathrm{C}$ at a rate of
3$\,^{\circ}\mathrm{C}$/h to grow single crystals. Single
crystals with shiny surfaces and typical dimensions of 5mm$\times5$mm$\times$0.2mm
were obtained. In the preparation process, the weighing, mixing,
grinding were conducted in a glove box under argon atmosphere with
the O$_{2}$ and H$_{2}$O below 0.1 PPM. The x-ray diffraction
(XRD) measurements were performed on a Bruker D8 Advanced
diffractometer with the Cu-K$_{\alpha}$ radiation. The DC
magnetization measurements were carried out with a SQUID-VSM-7T
(Quantum Design). The in-plane resistivity measurements were done
on a PPMS-16T (Quantum Design) with the standard four-probe
method.

\section{Results and Discussion}

\begin{figure}
\includegraphics[width=9cm]{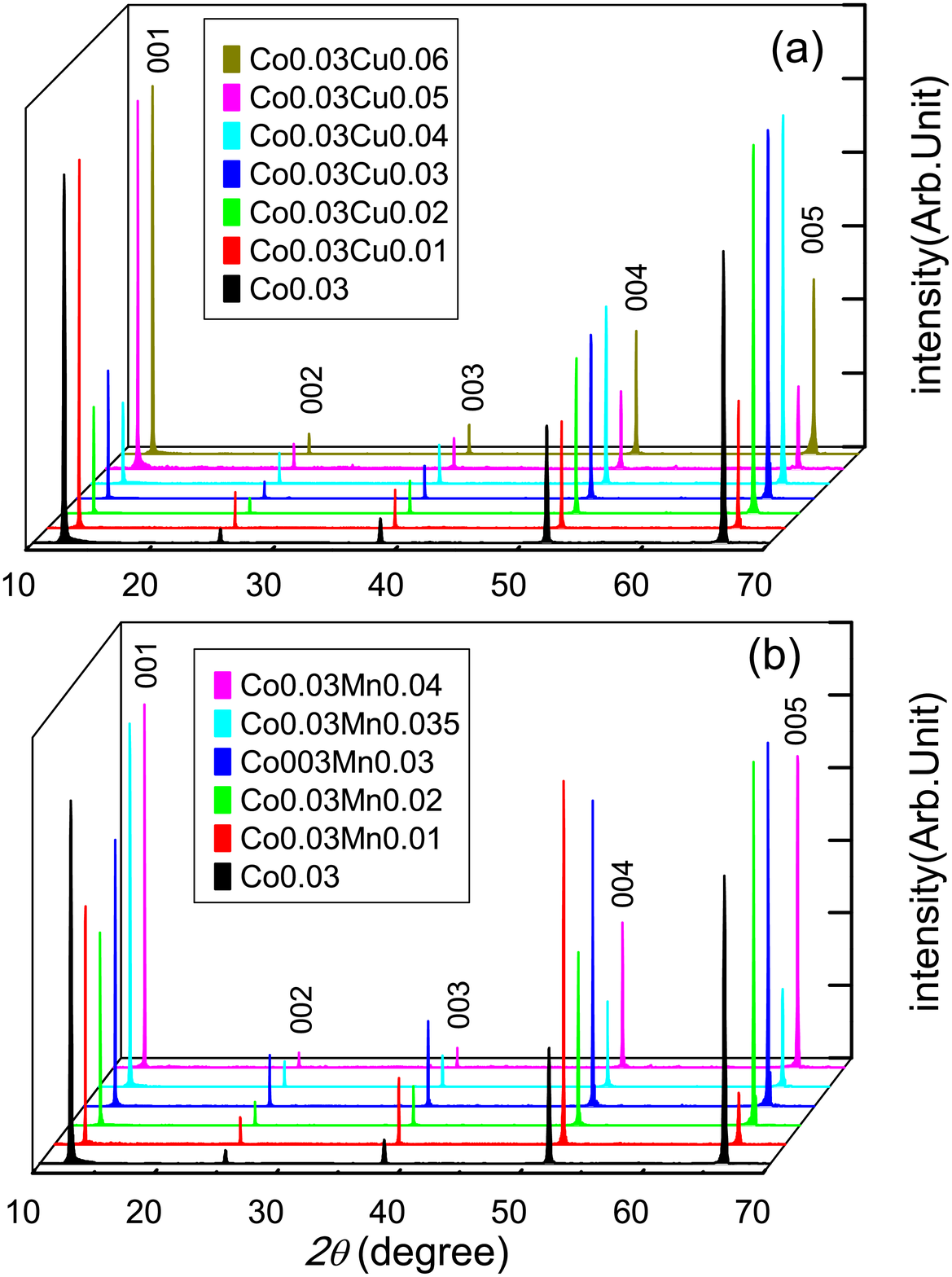}
\caption {(color online) The XRD patterns of (a) Na(Fe$_{0.97-x}$Co$_{0.03}$Cu$_{x}$)As,
and  (b) Na(Fe$_{0.97-x}$Co$_{0.03}$Mn$_{x}$)As single crystals.} \label{fig1}
\end{figure}

\begin{figure}
\includegraphics[width=9cm]{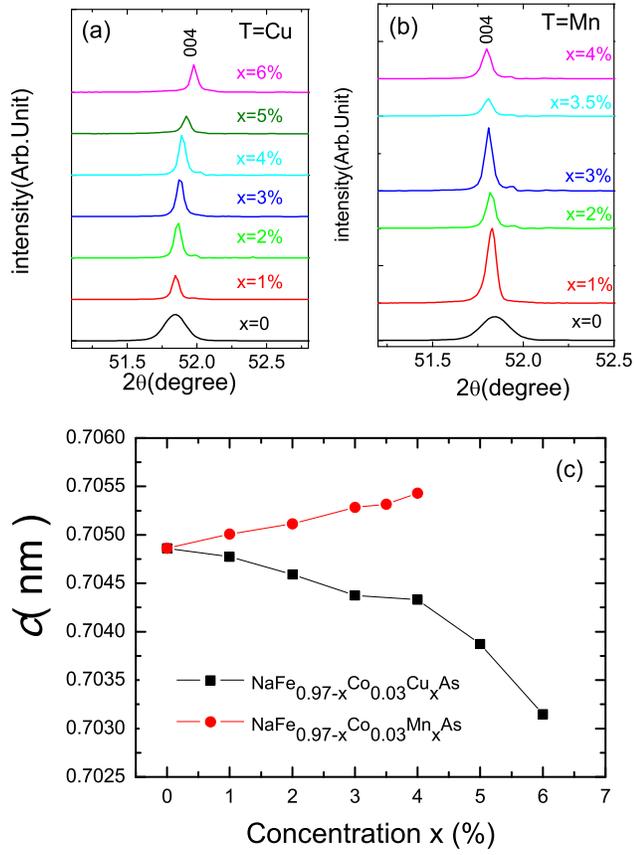}
\caption {(color online) (a) and (b) Peaks of the (004)
reflections of Na(Fe$_{0.97-x}$Co$_{0.03}$T$_{x}$)As (T = Cu and
Mn) single crystals. (c) The lattice parameter of c-axis plotted
as a function of the doping concentration $x$ for Cu and Mn doped
samples.} \label{fig2}
\end{figure}

\subsection{X-ray diffraction}

Fig.~1 shows the XRD patterns of the
Na(Fe$_{0.97-x}$Co$_{0.03}T_x$)As (T=Cu, Mn) single crystals. Only
(00$\emph{l}$) peaks can be observed, and all the diffraction
peaks show a full width at half maximum (FWHM) less than
0.05$^{\circ}$, indicating that the cleavage plane is the ab plane
and the high quality of the samples. As shown in Fig. 2(a) and
(b), peaks of the (004) reflection shift monotonously in 2$\theta$
with the increase of doping concentration, indicating that the
impurities were doped into the crystal lattice successfully. This conclusion is also supported by the monotonic increase of the residual resistivity versus doping level in both systems.
The lattice parameter of \emph{c}-axis is obtained and plotted as a
function of doping concentration $x$, as shown in Fig. 2(c). For
the pristine sample, the lattice parameter of \emph{c}-axis is 7.048
${\AA}$, which is consistent with the previously reported results
within experimental error.\cite{XHChen} We can see that, the \emph{c}-axis
lattice parameter slightly increases with the increase of Mn
doping concentration, while it decreases in the Cu-doped samples.
This behavior is similar to the results in other
reports.\cite{Frankovsky,JLi}

\begin{figure}
\includegraphics[width=9cm]{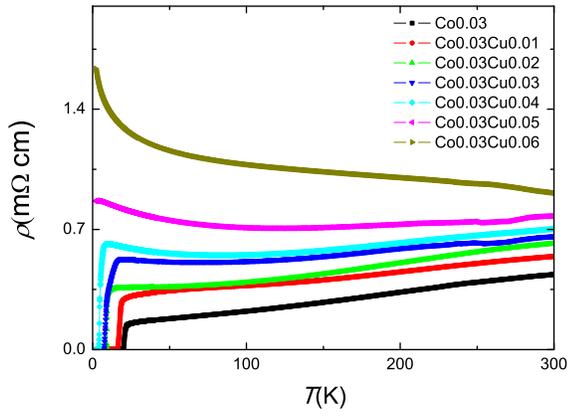}
\caption {(color online) Temperature dependence of the in-plane
resistivity of Na(Fe$_{0.97-x}$Co$_{0.03}$Cu$_x$)As. } \label{fig3}
\end{figure}

\begin{figure}
\includegraphics[width=9cm]{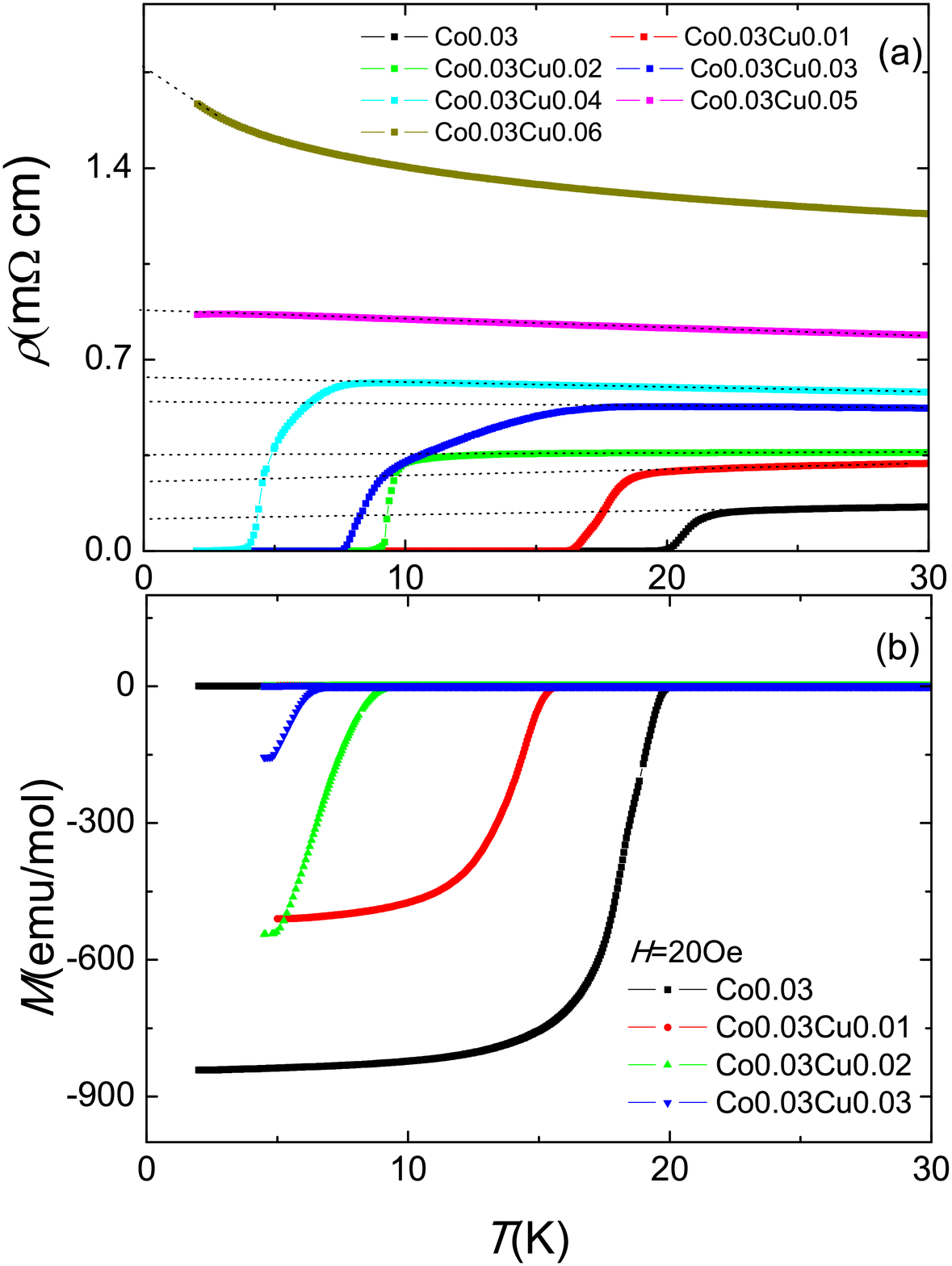}
\caption {(color online) (a) Temperature dependence of resistivity in
low temperature region for Na(Fe$_{0.97-x}$Co$_{0.03}$Cu$_x$)As.
The dashed lines represent the linear extrapolations of the normal
state data to zero temperature. (b) DC magnetization of
Na(Fe$_{0.97-x}$Co$_{0.03}$Cu$_x$)As.} \label{fig4}
\end{figure}

\begin{figure}
\includegraphics[width=9cm]{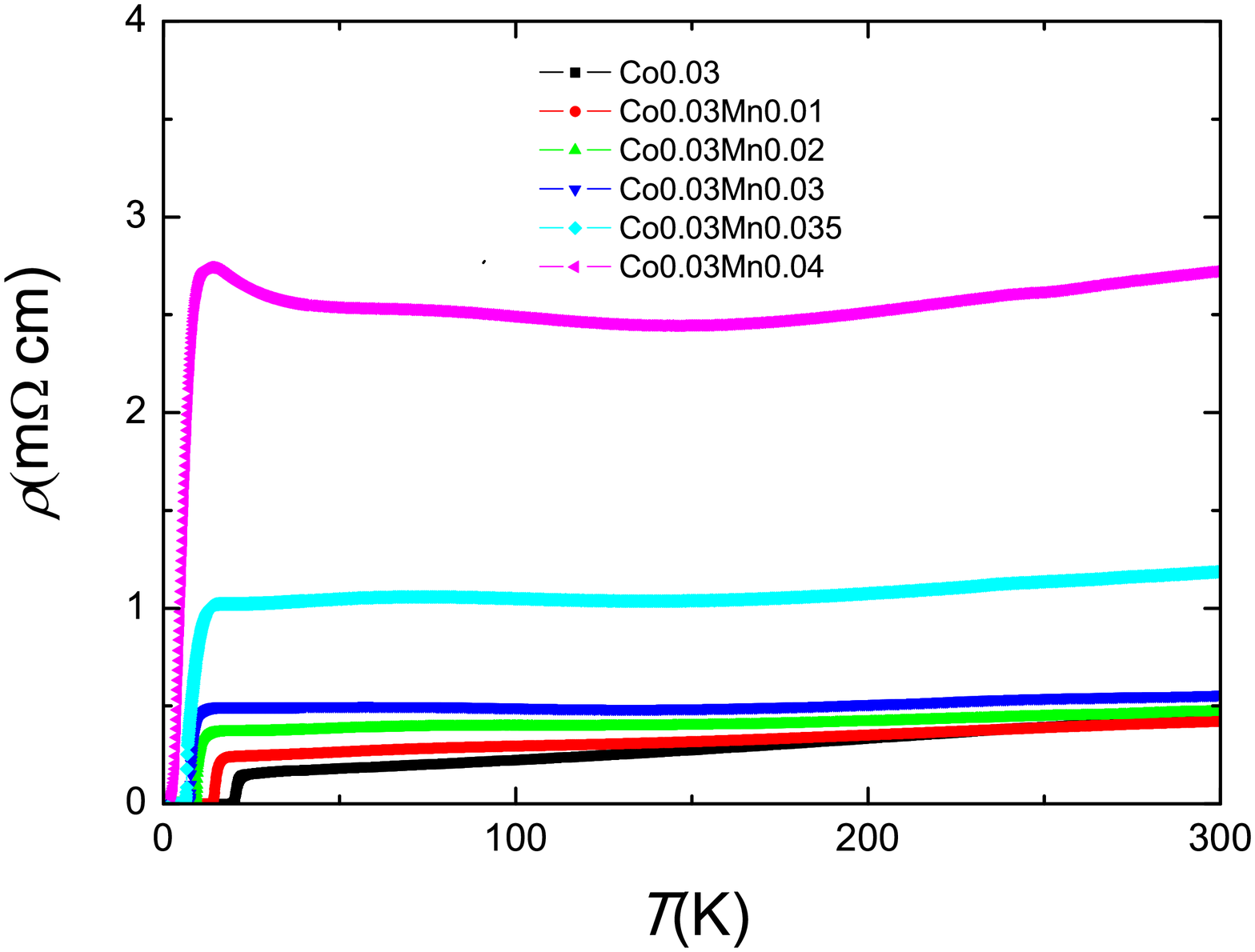}
\caption {(color online) Temperature dependence of the in-plane
resistivity of Na(Fe$_{0.97-x}$Co$_{0.03}$Mn$_x$)As.} \label{fig5}
\end{figure}

\begin{figure}
\includegraphics[width=9cm]{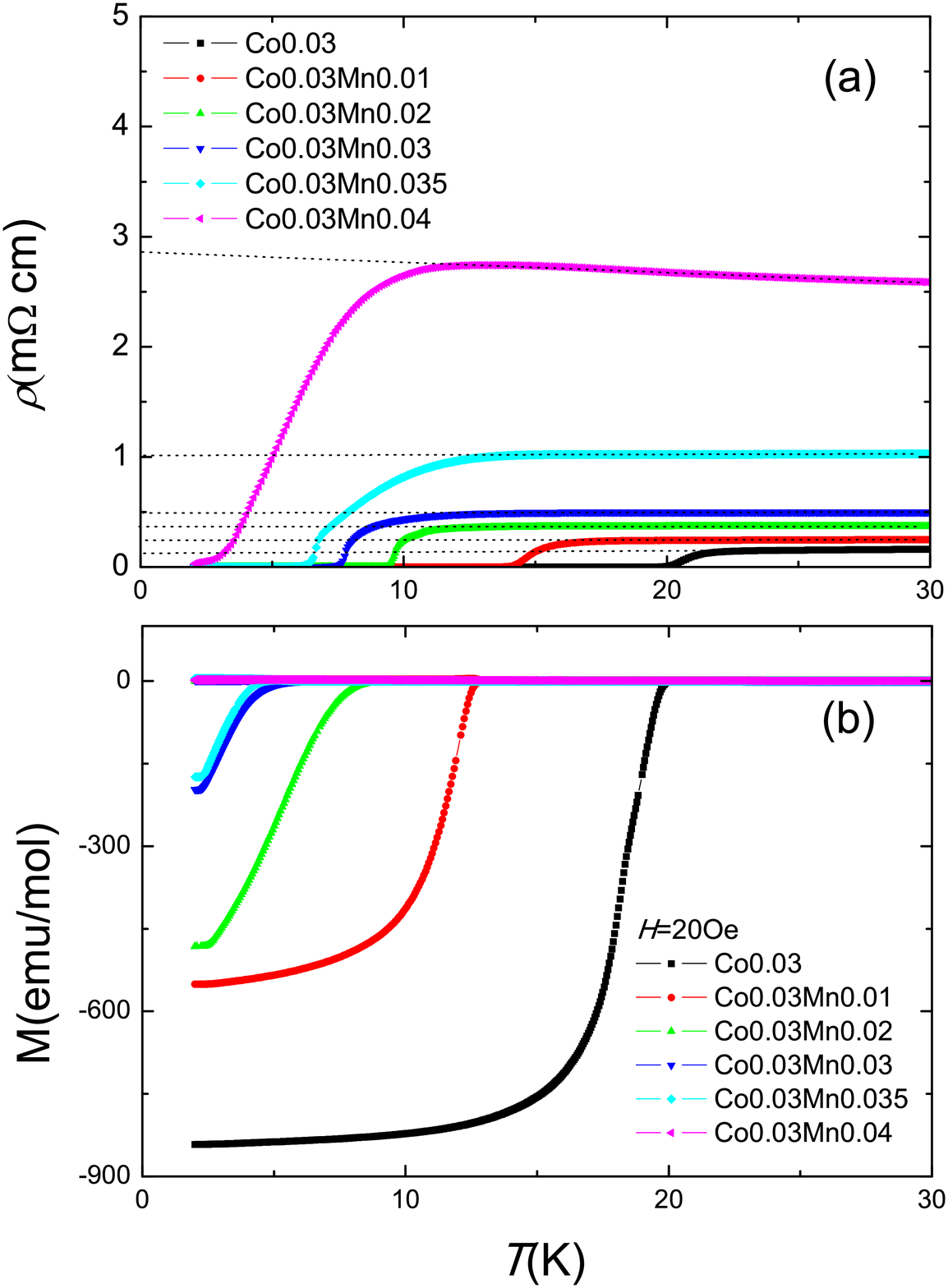}
\caption {(color online) (a) The resistivity curve in low temperature
region for Na(Fe$_{0.97-x}$Co$_{0.03}$Mn$_x$)As.
The dashed lines represent the linear extrapolation of the normal state data
to zero temperature. (b) DC magnetization of a
Na(Fe$_{0.97-x}$Co$_{0.03}$Mn$_x$)As single crystal. }
\label{fig6}
\end{figure}

\subsection{Magnetization and resistivity measurements}

In Fig. 3, we present the temperature dependence of the in-plane
resistivity from 2 K to 300 K for
Na(Fe$_{0.97-x}$Co$_{0.03}$Cu$_x$)As single crystals. Obviously,
\emph{T}$_{c}$ goes down and the residual resistivity goes up with the
increase of Cu concentration. In the low doping region, the
resistivity decreases upon cooling, followed by a superconducting
transition. For the highly doped sample, an upturn is observed
at low temperatures. It is interesting to note that, for the sample $x=0.03$, the transition seems to be broad, an initial drop of resistivity starts at about 16 K, which is followed by a sharper drop at about 8-9 K. This may be induced by a chemical segregation. But the strange point is that this broadened transition occurs in most of the measured curves of this doping level, even from the samples of different batches. We notice that, actually the normal state starts to show a low-$T$ upturn at this doping point. The low-\emph{T} upturn gets stronger with the
increase of Cu content, which may be induced by a stronger localization effect. A strong semiconducting like
temperature dependence of resistivity is observed on the sample
with x=0.05-0.06. Interestingly, it is found that \emph{T}$_{c}$ is suppressed to zero at the threshold of
strong semiconducting behavior. For the sample with x=0.05, a downward trend of resistivity
is observed at about 2.5 K. So the doping level x=0.05 is regarded as the critical doping
concentration which suppresses \emph{T}$_{c}$ to zero. We define the
residual resistivity $\rho$$_{0}$ by extrapolating normal
state data in a linear way in the low temperature region to zero temperature, as shown in Fig. 4(a). One can see,
\emph{T}$_{c}$ decreases monotonously with the increase of $\rho$$_{0}$,
and it is suppressed to zero at a residual resistivity of 0.87 m$\Omega$ cm.
Fig. 4(b) shows the temperature dependence of DC magnetization taken at 20 Oe
after the zero-field-cooling (ZFC) and field-cooling (FC)
procedure for the superconducting
Na(Fe$_{0.97-x}$Co$_{0.03}$Cu$_x$)As single crystals. In this
study, \emph{T}$_{c}$ is defined on the magnetization curve using the crossing point of the normal state background line and the extrapolated linear line of the steep transition, which
is nearly consistent with the point where resistivity reaches zero. For the
pristine sample, \emph{T}$_{c}$ reaches about 20.5 K.

The temperature dependence of the in-plane resistivity and DC magnetization of
Na(Fe$_{0.97-x}$Co$_{0.03}$Mn$_x$)As are shown in Fig. 5 and
Fig. 6 (b), respectively. Same as the Cu-doped samples, the residual resistivity
is determined by a linear extrapolation of normal state data to zero
temperature, as shown in Fig. 6(a). It is clear that
the residual resistivity increases with increasing the doping level,
and consequently \emph{T}$_{c}$ is also suppressed monotonically. Interestingly, a similar broadening effect of resistivity as the case of doping Cu occurs at about x=0.035, which seems to be some kind of intrinsic feature. Compared with Cu doped
samples, Mn doped ones show much weaker localization even to a very high doping, and the enhancement
of residual resistivity changes from a slow speed to a fast one.
Interestingly, superconductivity still exists in the sample with a residual resistivity even up 2.86 m$\Omega$ cm. One may argue that the Mn dopants may not be successfully doped into the system when the concentration is higher than a certain value. However, this argument cannot be supported by the doping dependence of the \emph{c}-axis lattice constant, as shown in Fig. 2(c). Moreover, the residual resistivity of the Mn doped samples increases also continuously, which again indicates a successful doping of Mn into the material. Clearly, the Mn impurity shows weaker
\emph{T}$_{c}$ suppression effect than Cu in the high doping regime. This remains to be an interesting and unresolved observation.

\begin{figure}
\includegraphics[width=9cm]{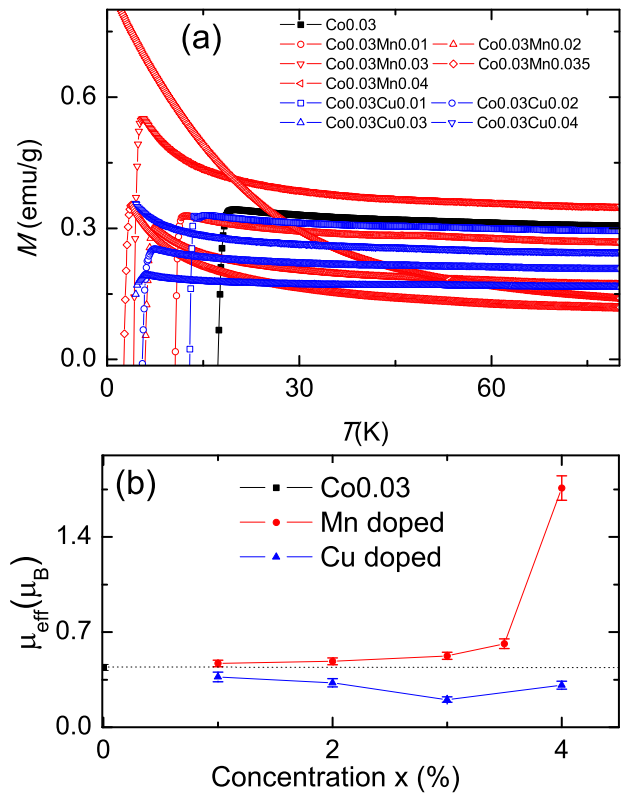}
\caption {(color online) (a) The magnetization measured at $\mu$$_{0}$H = 1 T on
the pristine sample (black squares), the Cu doped samples (blue symbols)
and the Mn doped samples (red symbols). One can see that the low
temperature upturn gets enhanced clearly by the Mn doping,
but not enhanced by the Cu doping.
(b) The average magnetic moment per Fe site calculated by the
Curie-Weiss law for the Cu and Mn doped samples. Clearly, doping
Mn induces strong averaged magnetic moments, while doping Cu seems
to weaken the averaged local magnetic moments.} \label{fig7}
\end{figure}

\begin{figure*}
\includegraphics[width=18cm]{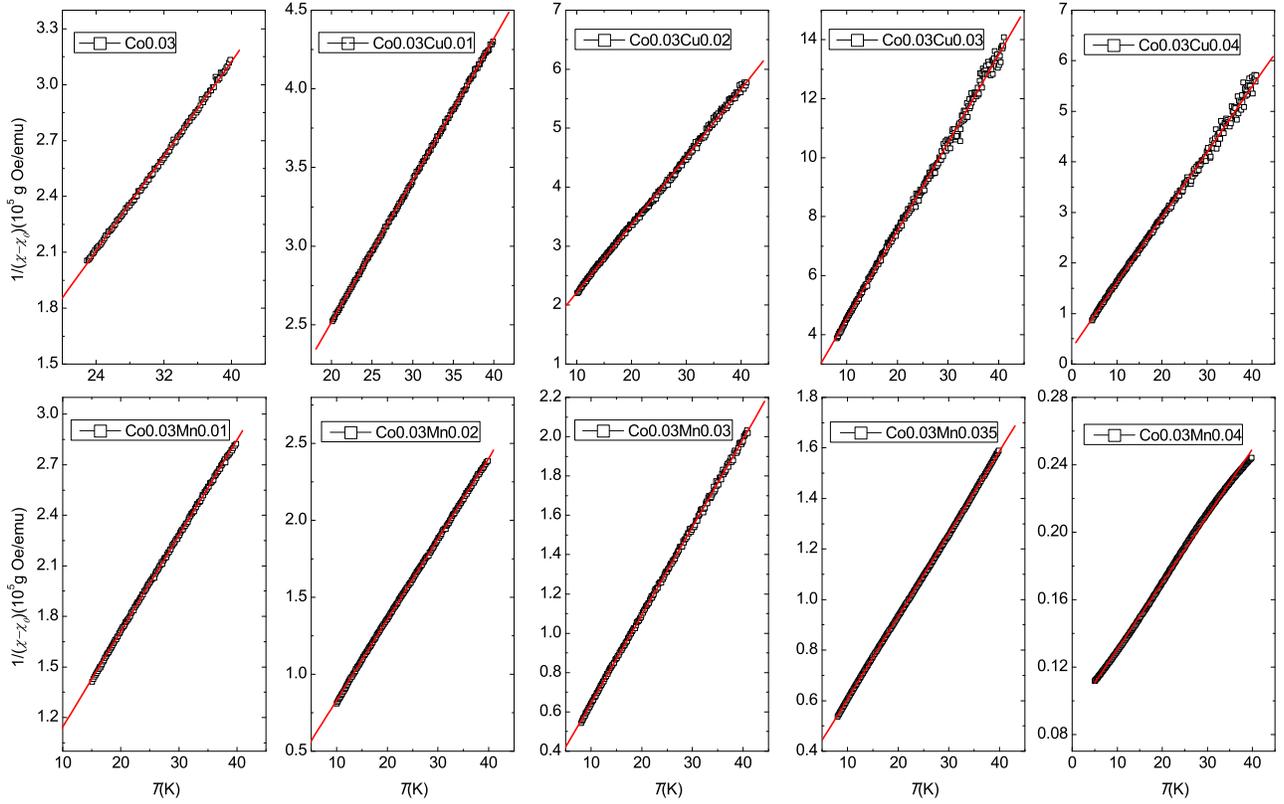}
\caption {(color online) Temperature dependence of
1/($\chi$-$\chi$$_{0}$) for Na(Fe$_{0.97}$Co$_{0.03}$)As and
Na(Fe$_{0.97-x}$Co$_{0.03}$T$_{x}$)As (T = Cu and Mn) single
crystals under 1 T with $\chi$ the DC magnetic susceptibility. By
adjusting $\chi_0$ we obtain a linear relation of
1/($\chi$-$\chi_{0}$) with temperature in the low temperature
limit. The red lines represent the linear fits of the data. The
slope gives 1/$C_0$ and the intercept provides the value of
T$_{\theta}$/$C_0$.} \label{fig8}
\end{figure*}

\begin{figure}
\includegraphics[width=9cm]{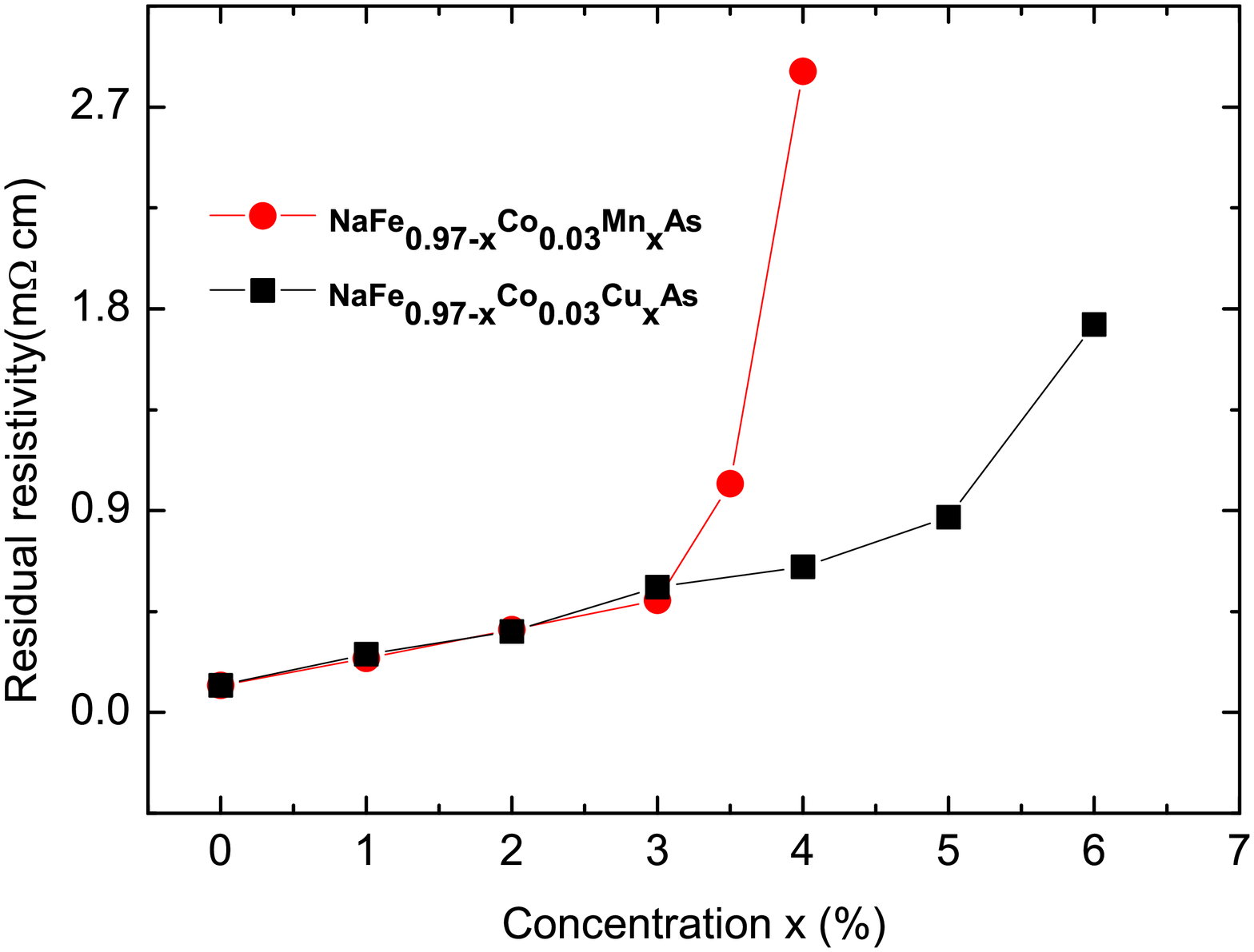}
\caption {(color online) Residual resistivity of
Na(Fe$_{0.97-x}$Co$_{0.03}$T$_x$)As (T=Cu, Mn) plotted as a
function of doping concentration x.} \label{fig9}
\end{figure}

\begin{figure}
\includegraphics[width=9cm]{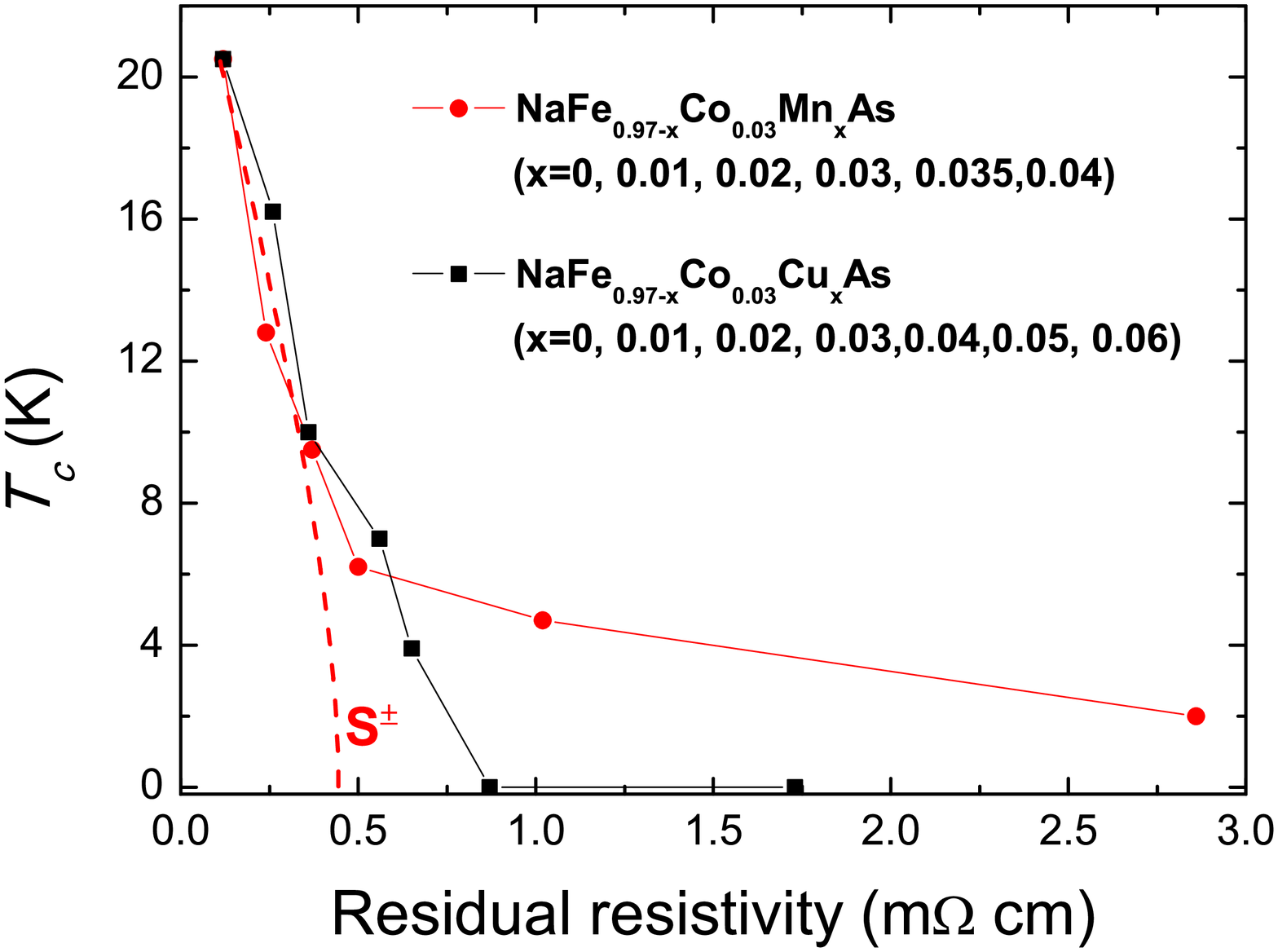}
\caption {(color online) Correlations between \emph{T}$_{c}$ and the residual resistivity for the investigated samples. The red dashed line
represents the relationship between \emph{T}$_{c}$ and residual resistivity
based on the S$^{\pm}$ scenario, which is calculated using the relation $\rho$$_{0}$=2$\pi$$\alpha$\emph{k}$_{B}$m$^{*}$\emph{R$_{H}$}\emph{T}$_{c0}$/z$\hbar$e+$\rho$$^{pri}_{0}$ and follows a universal Abrikosov-Gor'kov
formula. Here $\rho$$^{pri}_{0}$ represents the residual resistivity of the pristine sample.} \label{fig10}
\end{figure}

\subsection{DC magnetization and analysis}
In order to study the impurity scattering mechanism, it is crucial to determine whether
the impurity is magnetic or nonmagnetic. To evaluate the magnetic moments induced by Cu and
Mn dopants, we have done the magnetization measurements under high
magnetic fields. The raw data of magnetization measurements at 1 T
are shown in Fig. 7(a). The clear divergence of the
magnetic susceptibility at low temperatures can be understood as
the existence of some local magnetic moments. One can see that the
low temperature upturn gets enhanced clearly by the Mn doping,
but not enhanced by the Cu doping, indicating that Mn is the magnetic impurity,
while Cu behaves as non- or very weak magnetic impurity. To further investigate the
magnetic moments induced by Cu and Mn dopants, we assume that the
low temperature magnetization can be written in the Curie-Weiss
law,
\begin{equation}
\chi=M/H=\chi_{0}+C_0/(T+T_{\theta}),
\end{equation}
where
$C_0=\mu_{0}\mu^{2}_{eff}/3k_{B}$,
$\chi$$_{0}$ and $T_{\theta}$ are the fitting parameters,
$\mu$$_{eff}$ is the local magnetic moment per Fe
site. The first term $\chi$$_{0}$ comes from the Pauli
paramagnetism of the conduction electrons, which is related to the
density of states (DOS) at the Fermi energy. The second term
$C_0/(T+T_{\theta})$ is given by the
local magnetic moments of the ions at the Fe sites (including dopants like Co, Cu, and
Mn). The fitting process is not straightforward, for a precise evaluation on the local magnetic moments, we adjust the $\chi$$_{0}$ value to get a
linear function of 1/($\chi$-$\chi$$_{0}$) versus $T$ in the low
temperature limit. Then we fit the data with a linear function, as shown in Fig.8. The slope of the linear line gives $1/C_0$, and the intercept delivers the value
of $T_{\theta}/C_0$. Once $C_0$ is obtained, we can
get the average magnetic moment of a single Fe site (including the
contribution of Fe and the dopants). The results are shown in
Fig. 7(b). Clearly, doping Mn induces strong local magnetic
moments, while doping Cu seems to even weaken the average local
moments. These facts suggest that Mn ions here play as a role of
magnetic impurities, while Cu dopants act as the non- or weak
magnetic impurities. One possible picture to interpret this is
that the Cu dopant may have a full shell of d$^{10}$ configuration
with the ionic state of Cu$^{1+}$ as predicted by the theoretical
calculations.\cite{Chadov,Wadati} Similar results are observed in
Cu-doped Fe$_{1+y}$Te$_{0.6}$Se$_{0.4}$.\cite{ZTZhang}

\subsection{Discussions}
The residual resistivity of Na(Fe$_{0.97-x}$Co$_{0.03}$T$_x$)As (T=Cu, Mn)
plotted as a function of doping concentration x is shown in Fig. 9. One
can see that both doping Cu and Mn can enhance the residual resistivity.
In the low doping region (x=1\%, 2\%, 3\%), the residual resistivity increases
with the same ratio of 0.18 m$\Omega$ cm/\% for both Cu- and Mn-doped samples.
However, the residual resistivity increases rapidly for highly Mn
doped samples. Surprisingly, superconductivity  still exists even up
to a residual resistivity of 2.86 m$\Omega$ cm for the case of Mn doping. Similar phenomenon
is observed in Co-doped Fe$_{1+y}$Te$_{0.6}$Se$_{0.4}$
superconductor,\cite{ZTZhang} where superconductivity maintains
even up to a residual resistivity of 6 m$\Omega$ cm.  We must emphasize that the Mn elements have been doped into the Fe sites without doubt, because the lattice constant changes monotonously, and the high residual resistivity increases monotonously with the doping level of Mn up to 4\%. For a simple S$^\pm$ pairing model, this is very difficult to understand. A further in-depth understanding is highly desired.

In above we have discussed the influence of doping Cu and Mn on the lattice
parameter, magnetization and resistivity. We have also
discovered that Mn can enhance the local magnetic moments
greatly, whereas Cu behaves as a nonmagnetic impurity. Based on these
results, in the following, we focus on the discussion of pair-breaking mechanism
for Na(Fe$_{0.97-x}$Co$_{0.03}$T$_x$)As (T=Cu, Mn)
superconductors. According to the S$^\pm$ scenario with equal gaps of opposite
signs on different Fermi surfaces, \emph{T}$_{c}$ is expected to
be markedly suppressed due to potential scattering by substituted
nonmagnetic impurities, and it obeys a universal Abrikosov-Gor'kov
formula,\cite{Chubukov} -$\ln$t=$\psi$(1/2+$\alpha$/2t)-$\psi$(1/2), where
t=\emph{T}$_{c}$/\emph{T}$_{c0}$ with $T_{c0}$ and $T_c$ the transition temperatures of the pristine and the doped samples respectively, $\psi$(x) is the di-gamma
function, $\alpha$ is the pair breaking parameter.\cite{Abrikosov} The Abrikosov-Gor'kov formula would lead to that \emph{T}$_{c}$ vanishes
at $\alpha^{theory}_{c}$=0.28. However, plenty of previous studies of the
impurity effect on the iron-based superconductor have revealed
that the critical value of $\alpha$ is much larger than 0.28, which means that the rate
of \emph{T}$_{c}$ suppression found previously is too small to be explained
by the S$^{\pm}$ scenario.\cite{PCheng2,JLi,JLi2,JLi3,Inabe,Kirshenbaum}
According to the theoretical caculations based on the five-orbital model,\cite{Onari} $\alpha$=z$\hbar$$\Gamma$/2$\pi$\emph{k}$_{B}$\emph{T}$_{c0}$, where z=m/m$^{*}$ is the
renormalization factor, $\Gamma$ is the scattering rate. We use the relation $\Gamma$=ne$^{2}$$\Delta$$\rho$$_{0}$/m$^{*}$=e$\Delta$$\rho$$_{0}$/m$^{*}$\emph{R$_{H}$}, where $n$ is the carrier density, \emph{R$ _{H}$} is the Hall coefficient, m$^{*}$ is
the effective mass. In this study, we use z=1/4.2 obtained from optical spectroscopy
experiment for Na-111 superconductor,\cite{ZPYin,WZHu}
\emph{R$ _{H}$}=-7$\times$10$^{-9}$ m$^{3}$/C is obtained from the transport measurements by
extrapolating \emph{R$ _{H}$} data to zero temperature, which is consistent with the result reported before.\cite{XHChen} The obtained critical value of $\alpha$ for \emph{T}$_{c}$ to vanish based on our experimental data of Cu-doped sample is about 0.64, which is still much larger than the theoretically expected value $\alpha_c$ = 0.28, but the gap between the experimental and the theoretical value becomes much smaller compared with the previous studies in the Ba122 system.\cite{JLi,JLi2,JLi3}

To further study the pair-breaking mechanism in
Na(Fe$_{0.97-x}$Co$_{0.03}$T$_x$)As (T=Cu, Mn) system, we
calculated the critical residual resistivity for \emph{T}$_{c}$ to vanish based on the S$^{\pm}$
scenario, which is proportional to pair breaking parameter $\alpha$. The relationship between \emph{T}$_{c}$ and $\alpha$ can be transformed into the relationship between \emph{T}$_{c}$ and residual resistivity by using the relation $\rho$$_{0}$=2$\pi$$\alpha$\emph{k}$_{B}$m$^{*}$\emph{R$_{H}$}\emph{T}$_{c0}$/z$\hbar$e+$\rho$$^{pri}_{0}$, here $\rho$$^{pri}_{0}$ represents the residual resistivity of the pristine sample. As shown in Fig. 10, \emph{T}$_{c}$ is plotted as a function of $\rho$$_{0}$.
The dashed line represents the relationship between \emph{T}$_{c}$ and residual
resistivity based on the S$^{\pm}$ scenario, which follows a universal
Abrikosov-Gor'kov formula.
One can see that, the critical residual resistivity
for \emph{T}$_{c}$ to vanish in the S$^\pm$ model is 0.44 m$\Omega$ cm.
As we can see, in the low doping region, our
data can be roughly fitted to this model, which means that the impurity effect in
Na(Fe$_{0.97}$Co$_{0.03}$)As induced either by Mn or Cu can be explained by the S$^{\pm}$ scenario in the low doping regime.
In the case of Mn doping, the behavior of suppression to \emph{T}$_{c}$ changes
from a fast speed to a slow one and keeps the superconductivity
even up to a residual resistivity of 2.86 m$\Omega$ cm, indicating that
the magnetic Mn impurities are even not as detrimental as the non-magnetic
Cu impurities to superconductivity in the high doping region. This is an interesting observation, further theoretical
and experimental efforts are expected to carry out why the superconductivity can be so robust with the Mn doping.

\section{Conclusion}

In summary, we studied the impurity effect
in single crystals of Na(Fe$_{0.97-x}$Co$_{0.03}$T$_x$)As
(T=Cu, Mn). Analysis of DC magnetization based on the Curie-Weiss law
indicates that, Mn doping gives magnetic impurity, whereas Cu dopants behave as non- or very weak magnetic impurities.
However, it is found that both doping Cu and Mn can enhance the residual resistivity and suppress the
superconductivity in the same rate in the low doping region, being
consistent with the prediction of the S$^{\pm}$ model. For the Cu-doped
system, the superconductivity is suppressed completely at a
residual resistivity of 0.87 m$\Omega$ cm, when a strong
localization effect is observed. However, in the case of Mn
doping, the behavior of suppression to \emph{T}$_{c}$ changes from a fast
speed to a slow one and superconductivity survives even up to a
residual resistivity of 2.86 m$\Omega$ cm. Clearly the magnetic
Mn impurities are even not as detrimental as the non-magnetic
Cu impurities to superconductivity in Na(Fe$_{0.97}$Co$_{0.03}$)As system in the high doping regime.

\begin{acknowledgments}
We appreciate the useful discussions with Peter Hirschfeld, Igor Mazin. This work is supported by the
Ministry of Science and Technology of China (973 Projects: No.
2011CBA001002, No. 2010CB923002, and No. 2012CB821403), the NSF of
China, NCET project and PAPD.
\end{acknowledgments}

$^{\star}$ hhwen@nju.edu.cn

\end{document}